\documentstyle[12pt]{article}
\begin{document}
\renewcommand{\thefootnote}{\fnsymbol{footnote}}
\begin{center}
{\large\bf THE SOURCES OF SUPERGRAVITY\\ }
\vspace{1cm}
Vyacheslav A. Soroka\footnote{E-mail:  vsoroka@kipt.kharkov.ua}
\vspace{1cm}\\
{\it Institute of Theoretical Physics}\\
{\it National Science Center}\\
{\it "Kharkov Institute of Physics and Technology"}\\
{\it 310108 Kharkov, Ukraine}\\
\end{center}
\vspace{1cm}
\begin{abstract}
Recollections on how the basic concepts and ingredients of
supergravity were formulated by Dmitrij V. Volkov and the present author in
1973-74.
\end{abstract}
\renewcommand{\thefootnote}{\arabic{footnote}}
\setcounter{footnote}0

\bigskip
When I was an undergraduate student at the Kharkov State University in
1965 my advisor Dmitrij Vasilyevich Volkov gave me a task to learn and to
study the problem of quantization of the spin 3/2 field interacting with
other fields.  There was a problem with constructing selfconsistent
theories for the interacting fields with high spins and to find the
methods for quantization of such theories. In particular, Johnson and
Sudarshan (1961) encountered the inconsistency under quantization of the
spin 3/2 field minimally interacting with the external electromagnetic
field.  At that time I was a starting physicist and did not know the
theories with high spins. So, I had to learn two formalisms for
description of the spin 3/2 field:  Bargman-Wigner and Rarita-Schwinger.
By the way, from the paper of Rarita and Schwinger (1941) I learned and
memorized for the future a remarkable fact that a free massless spin 3/2
field in their formalism possesses a gauge invariance with a spinor
parameter.  I also learned Schwinger's quantum dynamical principle within
the framework of which Johnson and Sudarshan had tried to quantize the
spin 3/2 field.

At that time we had no success in quantization of the spin 3/2 fields
interacting with other fields. As became clear later on, the problem
in that formulation had no solutions at all. But it was elucidated only
after the discovery of supergravity theory in which the problem was
naturally solved. So, a solution of the problem came to a deadlock and
I was forced to postpone it and switched my activity to the
derivation of sum rules following from the algebra of fields
which had been introduced by Kroll, Lee, Weinberg and Zumino (1967).
Since the algebra of fields is based on the gauge theory, I learned
the beautiful Yang-Mills theory, and the papers by Utiyama (1956) and
Kibble (1961) devoted to the gauge theories with different gauge
groups, including, in particular, the gauge space-time symmetries connected
with gravity.

Meanwhile Volkov (1969) simultaneously with Callan, Coleman, Wess and
Zumino (1969) developed a general method for constructing Lagrangians
for the Goldstone particles in the case of an arbitrary spontaneously
broken internal symmetry group. Later he extended this method to the
case of the spontaneously broken symmetries including the Poincar\'e
group as a subgroup (1971, 1973).

Then Gol'fand and Likhtman (1971) introduced a super-Poincar\'e group in
connection with a parity violation problem in quantum field theory.
Shortly, and independently of them, Volkov formulated the existence
problem for the fermionic Goldstone particles with spin 1/2. A successful
solution of this problem led Volkov together with Akulov to the discovery
of the extended super-Poincar\'e group (1972).

After that Volkov stated the idea of gauging the super-Poincar\'e group,
and suggest I try to realize it. I accepted the suggestion with great
enthusiasm, and we started. Then my earlier
experience concerning the formalisms for high spin fields and gauge
theories was very suitable and useful. We decided to consider the
spontaneously broken extended super-Poincar\'e gauge group in order to
study the
Higgs effect for the Goldstone particles with spin 1/2 which had been
recently introduced.  Under this investigation very interesting and
important features were revealed \cite{vs1,vs2} .
\begin{itemize}
\item
First of all, the real gauge fields with spin 3/2
were introduced as graviton superpartners. This fact was very unusual for
those times, because till then the gauge fields possessed only integer
spins.  The graviton superpartners with spin 3/2 were later called the
gravitino fields.
\item
Secondly, the extended super-Poincar\'e gauge group
gives a principle possibility for the nontrivial
unification of gravity with the interactions based on the internal
symmetries.
\item
Thirdly, the Higgs effect for the Goldstone particles with
spin 1/2, later called the super-Higgs effect, essentially differs from
the Higgs effect for the Goldstone particles with spin 0. It results in
not only the gravitinos becoming massive, but also the space-time
changes its own metric and topological properties: a nonzero cosmological
constant appears.
\end{itemize}

We have written down our action as a sum of five
invariant terms with arbitrary constants before them. These terms contain
respectively:  the Einstein--Cartan action for gravity, a kinetic term for
the gravitino fields, a mass term of the gravitinos, a cosmological
term and a term for the Yang--Mills fields with spin 1.

Thus, we see that the main notions and ingredients of supergravity were
formulated in our papers \cite{vs1,vs2}.

Soon after and independently Wess and Zumino (1974) introduced a
four-dimensional superconformal group generalizing the  group of
two-dimensional supersymmetric transformations
found earlier in the dual models by Ramond, Neveu, Schwarz, Gervais and
Sakita (1971).
Then Wess and Zumino proposed a supersymmetric model (1974) which appeared
to be renormalizable.

The main results of our works were used afterwards as starting points for
the development of supergravity in the works by Ferrara, Freedman, van
Nieuwenhuizen \cite{ffn} and Deser, Zumino \cite{dz}. In particular,
accepted in Ref. 4 transformation rules of the vierbein, used for
the graviton description, and gravitinos coincide with ours. Explicit use
of nonzero torsion, found in supersymmetric theories by us \cite{avs1} and
Zumino \cite{z}, also essential simplified the derivation of supergravity
in Ref. 4. Moreover, as Volkov explained \cite{v}, pure N = 1
supergravity \cite{ffn,dz} can be deduced from our theory \cite{vs1,vs2}
with a special choice of the constants of our invariant terms.

The notions of superspace, introduced by Akulov and Volkov (1972), and
superfields, proposed by Salam and Strathdee (1974), were used by
Arnowitt, Nath and Zumino in 1975 to initiate a superfield approach to
supergravity.  In order to overcome the drawbacks they found at their
first step in this direction, we \cite{avs1} undertook,
simultaneously with Zumino \cite{z}, the development of the superspace
approach to supergravity. By generalizing to superspace Cartan's
methods in differential geometry, we revealed a nonzero torsion of
superspace in the supersymmetric theories and found that the true
homogeneous holonomy group for the superspace in supergravity is the
Lorentz group. These two points are part of the basis of any version of
supergravity.

Let me also mention our construction in the superfield formulation
of a version of N = 1 supergravity in the linear approximation with the
so-called new minimal set of auxiliary fields \cite{avs2}, which we
had found earlier than the old minimal set that was obtained by Ferrara,
van Nieuwenhuizen, Stelle and West (1978).

So, I recalled the steps in supergravity which had been performed in
our works in 1973-1977.

As the complete list of papers on the theme of my reminiscence is very huge
and essentially exceeds this note, I refer only to our works and to those
very closely related to them. A very extensive list of references
concerning the development of supergravity can be found in Ref. 9.
This list contains also our papers \cite{vs1,vs2,avs1,avs2}, however, the
references to the latter are omitted in the text of this review.

To finish my note, I want to cite a very surprising (for me)
appraisal of supersymmetry given by Yuri Abramovich Gol'fand
during the Conference "Supersymmetry-85" at the Kharkov State University in
1985.  He said that supersymmetry did not justify his hopes to find a
generalization of the Poincar\'e group such that every its representation
includes particles of different masses. So, Gol'fand and Likhtman
had missed their aim, but had instead found supersymmetry, every
representation of which contains fields of different spins. Maybe,
somebody will succeed in their original aim too, or maybe both
aims at once.  Time will show.

In conclusion I would like to thank the organizers of this Symposium for
the opportunity to present my recollections concerning the creation of the
main ideas of supergravity at its early stages, which are usually not
illuminated.

\end{document}